\documentclass[12pt,preprint]{aastex}

\begin{document}

\title{\bf Host Galaxy Evolution in Radio-Loud AGN}

\author{Matthew O'Dowd}
\affil{School of Physics, University of Melbourne, 
Parkville, Victoria 3010, Australia; {\em modowd@physics.unimelb.edu.au}}
\author{C. Megan Urry}
\affil{Department of Physics and Yale Center for Astronomy
and Astrophysics, P.O. Box 208121, New Haven, CT 06520-8121, USA; {\em meg.urry@yale.edu}}

\begin{abstract}
We investigate the luminosity evolution of the host galaxies
of radio-loud AGN through Hubble Space Telescope imaging of 72 BL~Lac 
objects, including new STIS imaging of nine $z > 0.6$ BL~Lacs. 
With their intrinsically low accretion rates and their
strongly beamed jets, BL~Lacs provide a unique opportunity to probe host galaxy
evolution independent of the biases and ambiguities implicit in
quasar studies. We find that the host galaxies of BL~Lacs evolve 
strongly, consistent with passive evolution from a period of active
star formation in the range 
$0.5 \lesssim z \lesssim 2.5$, and inconsistent with either passive
evolution from a high formation redshift or a
non-evolving population. This evolution is broadly consistent with
that observed in the hosts of other radio-loud AGN, and inconsistent
with the flatter luminosity evolution of quiescent early types and
radio-quiet hosts. This indicates that active star formation, and hence
galaxy interactions, are associated with the formation for radio-loud
AGN, and that these host galaxies preferentially accrete less
material after their formation epoch than galaxies without 
powerful radio jets. We discuss possible explanations
for the link between merger history and the incidence of a radio jet.

\end{abstract}


\section{Introduction}\label{intro}

From Hubble Space Telescope (HST) studies of local galaxies, it
appears that nearly all contain supermassive black holes (SMBHs)
\citep{Joseph, Bower, Sarzi, Barth, Gebhardt}. 
Combined with the predictions from hierarchical galaxy formation,
this suggests that the formation and evolution of galaxies and SMBHs 
might be closely linked
\citep{Silk, HaehneltR, Kauffmann, Franceschini, Merrifield, Wang, diMatteo}. 
Circumstantial evidence also points to a direct link between the
evolution of normal galaxies with the AGN phenomenon --- the 
co-moving densities of both rise rapidly
from today back to redshift $\sim 2$ \citep{Schmidt, Boyle1,
Shaver1, Lilly1, Driver, Cowie, Madau, Connolly, Cowie2,
Hasinger}. Certainly it is
plausible that AGN luminosity affects host galaxy properties,
particularly star formation rates 
(e.g. de Young 1989; Rees 1989; Whittle 1992; Gonzalez Delgado 1995; 
Barthel 2001; Zirm, Dickinson \& Dey 2003), 
and a number of studies have found a dependence of maximum AGN power on host
galaxy luminosity \citep{Smith2, Veron-Cetty, Hutchings2, Hooper,
McLeod, Schade, Dunlop}.

It may be that most
galaxies pass through one or more active phases at some point in their
development \citep{Cavaliere}. If so, then comparing the luminosity
evolution of AGN host galaxies to that of normal galaxies allows us to
study the link between AGN activity and galaxy evolution.
If galaxy interactions are an important trigger of AGN activity
(e.g. Hutchings \& Campbell 1983; Heckman {\it et al.} 1984; Yee 1987; Heisler
1991; Hutchings \& Neff 1992; Canalizo \& Stockton 2001)
then there is a link between AGN activity and the 
assembly of elliptical galaxies, which should be reflected in
the luminosity evolution of AGN host galaxies \citep{Franceschini}.

Work on host galaxy evolution to date has largely been restricted to
high-power sources due to the difficulty in finding low-power AGN at
sufficiently high redshift for the evolution of galaxies to become
significant. The host galaxies of powerful radio-loud AGN at high
redshift appear to be brighter than general radio-loud host galaxies
observed at low redshift, consistent with passively evolving 
stellar populations \citep{Aragon-Salamanca, Lilly4, Kukula, McLure1,
  Lehnert}.
The same is not observed of brightest cluster galaxies (BCGs),
 which exhibit flat luminosity evolution over the
redshift range 0 to 1 \citep{Aragon-Salamanca}, and so are likely to be gaining 
in luminous mass. This is also tentatively the case for other early-type
galaxies \citep{Stanford, Lilly1, Kauffmann96, Bell}.
The host galaxies of radio-quiet AGN seem to exhibit similar flat
or negative luminosity evolution, and appear to be gaining mass
between redshift $z\sim 2$ and the present \citep{Kukula, Rix, Ridgway2}.

However, various biases make it impossible to confidently 
ascribe the observed evolution to the host galaxies themselves. 
First, it appears that maximum AGN power
increases with host mass \citep{Smith2, Veron-Cetty, Hutchings2,
  Hooper, Schade, McLeod, Scarpa3, ODowd, Dunlop}. This may induce
apparent redshift evolution in flux-limited samples.
Second, in powerful AGN, scattered light from the nucleus and
extended, ionized gas contaminates host galaxy
light. In radio galaxies, nuclear emission may be mistaken 
for host galaxy light \citep{DunlopP}.
This contamination increases with AGN power, 
which may again produce a spurious
host luminosity-redshift trend.

BL~Lac objects are
intrinsically low-power AGN, with much lower accretion rates than
quasars \citep{ODowd} or powerful radio galaxies, yet
the relativistic beaming of their jet emission means that they can be
detected in significant numbers to $z\sim 1$, at which point galaxy
luminosity evolution becomes measurable. Fortuitously, 
BL~Lac objects overcome many of the potential biases outlined above.

First, the apparent luminosity of BL~Lac objects is dominated by
magnified emission from their relativistically beamed aligned jets
\citep{UrryP}. This magnification is in 
turn dominated by the angle of the jet to the line of sight. 
Variation in jet angle washes out any trend between the power of the nucleus
and the luminosity of the host galaxy (supported by
absence of any correlation between host luminosity and beamed nuclear luminosity;
Urry {\it et al.} 2000.)
Second, the host galaxies of BL~Lac objects are less likely to suffer from
contamination by scattered light from their low-power
nuclei, nor do they tend to exhibit 
the luminous, extended, ionized gas observed in more powerful AGN.

Due to the $(1+z)^4$ surface brightness dimming of the hosts
(compared to $(1+z)^3$ dimming of the nucleus), the Hubble Space 
Telescope (HST) is needed to reliably determine BL~Lac host galaxy 
parameters beyond $z\sim0.6$. In this paper, we present new HST \footnote{Based 
on observations with the NASA/ESA Hubble Space Telescope, 
obtained at the Space Telescope Science Institute, which is
operated by the Association of Universities for Research in
Astronomy, Inc. under NASA contract No. NAS5-26555.} 
STIS images of high-redshift BL~Lac objects. Combined with 
earlier $z<0.6$ HST WFPC2 observations, we measure the evolution of
radio-loud AGN host galaxies, free from many of the biases implicit in
previous studies. The observations and data analysis are described in
\S\ref{obs}. Results on the host galaxy properties are in 
\S\ref{properties}, analysis and results for the host galaxy luminosity
evolution are presented in \S\ref{evolution}, and the discussion
of these results in \S\ref{discussion}. Conclusions are
summarised in \S\ref{conclusions}. 

\section{Observations and Data Analysis}\label{obs}

\subsection{The Sample}

The new STIS observations are high-redshift BL~Lac objects selected from
the cycle 6 HST snapshot survey of \citet{Urry}. 
The original 132 proposed snapshot
targets were chosen from six flux-limited samples. These were essentially
all known BL~Lac objects in complete samples in 1995. The 110
BL~Lacs finally imaged in the snapshot program were chosen at random from the larger sample,
and are an unbiased subset.
The sample chosen for STIS imaging comprised the 9 least optically luminous BL~Lacs with 
$z > 0.6$, spanning a range at least to $z\sim 1$. 
As observed nuclear brightness is dominated by variation in jet 
angle, this luminosity requirement does not introduce significant bias, 
but does dramatically increase our chances of detecting the host galaxy.
Table~\ref{targlist} gives the final target list and exposure information, with the magnitudes
and host galaxy limits derived from the WFPC2 snapshot images.

These targets were imaged with the STIS camera aboard HST, using the
F28$\times$50LP longpass filter.
This configuration maximizes
sensitivity to the host galaxy, sampling from 5,500\AA~ to 10,000\AA. 
The absence in BL~Lac objects of a big blue bump or extended line
emission, as seen in quasars, makes it possible to use such a broad filter.

\subsection{Data Reduction and Calibration}

Four to five exposures were taken for each target, dithered in a
square with edge length 5 pixels. In each case, one exposure was short
enough to ensure that the nucleus was not saturated.
The data were reduced using the STIS package in IRAF's STSDAS
suite. The bias and dark frames were subtracted and the images flat-fielded
using the routine STIS reduction process. CRREJ was used to
perform cosmic ray rejection.

Most of the long exposures resulted in over-exposure and bleeding of
the BL~Lac nuclei. In each of these cases the affected pixels in the
combined image were flagged and replaced with the pixels in the
shorter exposure, normalised according to their exposure times.

\subsection{Sky Subtraction}\label{sky}

Accurate determination of the background level and its uncertainty is
critical in this analysis, as an incorrect measurement can easily lead
to the false detection of host galaxies.
The background was taken to be the median in the region of each
combined imaged which was unaffected by any visible source. Such
sources were masked well beyond their visible extent for determination
of the background.

The error in the background value was calculated from two contributions: the
pixel-to-pixel Poissonian shot noise ($\sigma_{Poisson}$), and the
large-scale variation in the background level ($\sigma_{large-scale}$). 
The latter was taken to be the maximum of: standard deviation in the
medians of a grid of 20$\times$20 pixel squares, and the standard deviation
in the medians of the four quadrants of the image. The final
error in the subtracted sky value was then: 

\begin{center}
\begin{math}
\sigma_{sky} = \sqrt{\sigma_{Poisson}^2 + \sigma_{large-scale}^2}
\end{math}
\end{center}

Figure~\ref{rawpics} shows the central 300$\times$300 pixels of the
combined, sky-subtracted images. Also shown are the contour plots of
the regions surrounding the targets, with north indicated by the arrow
head and east indicated by the arrow tail.

\subsection{Modelling the STIS Point Spread Function}\label{psf}

Perhaps the most critical step in extracting the properties of the host
galaxies is the accurate modelling of the Point Spread Function (PSF). 
Unlike ground-based PSFs, the HST PSF is relatively stable with
time, and so synchronous PSF observations are not essential. However,
the PSF does vary significantly across the surface of the chip, and 
this must be taken into account. In all images the target 
was within 20 pixels of the centre, so such spatial variations
are minimized.

Due to deficiencies in the Tiny Tim model of the STIS PSF (it fails to
account for extended, scattered light, the STIS 'ghost loop' and an
asymmetric bulge near the PSF's centre), we elected to model the PSF
with archival images of well-centred stars taken with STIS in the
F28$\times$50LP filter. 
Highly over-exposed stellar images were used to model the PSF wings,
while unsaturated images were used to model the central regions. 
Figure~\ref{stispsf} shows the final composite stellar PSF.
The 'ghost loop', a reflection feature in the STIS PSF, was found to
vary significantly even with small positional offsets, and so we
simply mask this feature in all of the analysis to follow.

Given the potential inaccuracies in the STIS PSF extreme care was
taken to account for its uncertainties. Deviations in the PSF 
arise from spatial variations, time variations, and spectral
variations. The stars that were averaged to make our composite PSF
have different spectral types, were taken at different times, and are
offset from each other spatially (although all are still relatively
close to the centre). Thus, the deviations observed among these PSFs
should reflect the potential difference between the composite PSF and
our data. 
We define the uncertainty in each pixel of the composite PSF model to be
the maximum difference between that pixel value and the corresponding
pixels in the component stellar images. This gave a conservative
uncertainty map which was used in the two-dimensional host galaxy
fitting. 

\subsection{Host Galaxy Fitting}\label{fitting}

Host galaxy properties were measured using two-dimensional model
fitting. The models comprised the
stellar PSF (\S.\ref{psf}) centred on an analytic galaxy 
model. The host galaxies were modelled with the
following profile types: a de~Vaucouleurs $r^{1/4}$ law, an exponential
disk $r^{-1}$ law, and a composite de~Vaucouleurs plus exponential
disk model. 
The free parameters fitted for each host galaxy profile type were: PSF
normalization, host galaxy normalization (two normalizations for the
composite model), host galaxy characteristic radius (again, two
parameters for the composite model), ellipticity, and position angle.

Certain parameters were limited or fixed, as the quality of the data
did not justify more detailed fitting. The point source was fixed
at the centre of the galaxy, as this is observed to be the case all in low 
redshift BL~Lacs~\citep{Urry}. The galaxy models were limited to the three
profile types described, as we are primarily interested in distinguishing
between disk- and bulge-dominated morphologies. Modelling more detailed 
profiles (such as via the Sersic index) is beyond the scope of these data.

Before fitting, images were carefully masked to
eliminate all irregular features such as close companions and tidal
structure, along with several pixels beyond the visible edges of these
features. The centre-most few pixels of the nucleus are highly sensitive to
sampling errors. The values of these pixels will depend strongly on
exactly where in the pixel the point-like nucleus falls. If the model
PSF is offset even slightly from this (by as much as 0.05 of a pixel), 
large errors will occur in the fit, as these central pixels 
strongly weight our fitting statistic. To remedy this, we sum
the pixel values within a central circle with a 7-pixel diameter, and
treat this as a single data point. 
As a side-benefit, this process greatly reduces the potential error
due to an inaccurately modelled PSF.

In all cases but that of 2240--260, the fits indicated the presence host
galaxies. The residuals of 0820+225, 1144--379, 1249+174, 1422+580 and
1533+535 were smooth and either circularly or elliptically symmetric about the
nucleus. The residuals of 0138--097, 0235+164 and 1308+326 were less
convincing, so great care was 
taken to determine the confidence of the detections.

\subsubsection{Detection Criteria and Upper Limits}

To determine whether the apparent host galaxy detections were
significant, we fitted each image with the PSF alone, and compared the
$\chi^2$ values of the best PSF-only fit to the best PSF $+$ host
galaxy fit via the F-test. We required that the model including the
host galaxy be preferred over the PSF-only fit with 99\% confidence.

To make certain that we did not mistake improperly subtracted
background light for the host galaxy, we also perform the fits with
the background over-subtracted by $2 \sigma_{sky}$ (see \S\ref{sky}), and required 
that the model including the host galaxy was preferred over the
PSF-only fit with 95\% confidence in this case.

To ensure that none of the detected host galaxies were simply
artifacts of a poorly-modelled PSF, we redid the fits using each of
the individual stellar images as our PSF model instead of the combined
stellar PSF, and required again that the
host galaxy detected with each fit was preferred over the PSF-only fit
with 99\% confidence. For each alternate PSF, we again require the
host galaxy be preferred at 95\% with the background over-subtracted.

After applying these requirements, it was found that five of the nine
host galaxy were detected with high confidence. These were the host
galaxies of 0820+225, 1144--379, 1249+174, 1422+580 and 1533+535. \
Of the remainder, 0138--097,
0235+164 and 1308+326 yielded host galaxies in the initial fit, but these did
not meet our confidence requirements when the background was
over-subtracted and/or the alternate PSFs used. 2240--260 did not show a
detectable host galaxy in any of the fits. 

Figure~\ref{profiles} shows the one-dimensional, azimuthally averaged
profiles of both the data and the best fits. Galaxy profiles are 
presented only in cases where the PSF~+~host galaxy model is preferred over
the PSF-only model according to the initial resolution criterion;
ie. for 0138--097 and 0235+164 and the exponential disk fit of 1308+326, 
where the model was rejected only after increasing the
sky background or after fitting with an alternate PSF, we still
present the best-fit PSF~+~galaxy profiles.

Figure~\ref{imagesub} shows the images with PSF 
subtracted. For the detections we use the best PSF normalization from the 
de~Vaucouleurs fit, and for the non-detections we use the best PSF-only fit.
For the detections we also show the images with best-fit PSF + 
de~Vaucouleurs model subtracted. 

As a last check, we take the ratio of light in a aperture of
radius 3.5 pixels around the centre to the light an annulus from 15 to 50 pixels, 
and compare to the same ratio for the PSF. In all five of the
confident detections, the BL~Lac image showed a significant excess of
light in the annulus, indicating the presence of host galaxies.
In the non-detections the excess was marginal.

For the non-detections, we determined the upper limit on the host
galaxy magnitude by finding the maximum amount of host galaxy which
could be added before the PSF + host galaxy fit became worse than the
PSF-only fit at the 99\% confidence level by an F-test. In these
cases, the galaxy model used was a de~Vaucouleurs model with zero
ellipticity and an effective radius of 10~kpc --- slightly higher than
the median found for the low redshift BL~Lac hosts 
of 8.5~kpc \citep{Urry}.
This is conservative, as it will give a slightly higher upper limit
than for the median $r_e$.

\subsubsection{Uncertainty Analysis and Resolving Morphologies}

The uncertainty in each fitted parameter was determined by mapping
$\chi^2$ around the best fit. To account for the
uncertainty in the background subtraction, the fits were performed,
and the $\chi^2$ volume was mapped with the background both under- and 
over-subtracted by $\sigma_{sky}$ (see \S\ref{sky}). The
uncertainty in each parameter was then the maximum displacement from
that parameter on the one-sigma hypersurface of the $\chi^2$ volume.

For a statistically significant preference of one profile type over
the other, we again employ the F-test, 
requiring that their minimum $\chi^2$ values
differ with 99\% confidence. In most cases, the de~Vaucouleurs model
was marginally preferred over the exponential disk model, however in only two
of these cases was the preference significant --- for 0820+225 and 
1422+580. In no cases was the exponential disk profile preferred over
the de~Vaucouleurs profile with high confidence.

As might be expected, the composite de~Vaucouleurs + exponential disk
model gave a better fit than either the de~Vaucouleurs or exponential 
disk model alone. Typically, the improvement in the fit was marginal
--- at less than 90\% confidence according to the F-test in all cases but
0820+225, for which the three-component model was preferred over the
best-fit two-component model with $\sim$90\% confidence. 

A number of the de~Vaucouleurs model fits gave unphysically large
effective radii, suggesting that the fit may be trying to
compensate for an extended, disky component. Restricting
the PSF + de~Vaucouleurs model to more physical scale
lengths --- requiring $r_e \le 20$~kpc --- then the composite galaxy 
model is preferred over the new best-fit PSF + de~Vaucouleurs model for
0820+225 at the 99\% confidence level. If we 
apply the same condition to the rest of the sample, both 1422+580 and
1533+535 prefer the composite model at $\sim95$\% confidence,
and 1144$-$379 at the $\sim90$\% level.

Figure~\ref{profiles} shows $\chi^2$ contours projected onto the
$m_{host}$--$r_e$ plane for the de~Vaucouleurs and exponential fits of
the resolved sources. 

\subsection{K Corrections and Filter Conversions}

Cousin's R band absolute magnitudes are presented for purposes of
comparison with the low redshift BL~Lacs in \citet{Urry}. 
These corrections were performed by assuming SEDs for both the host
galaxies and the nuclei. The models of \citet{Bruzual} were used for
the host galaxy SEDs, calibrated to the galaxy colours reported in
\citet{Fukugita}: a Hubble-type Ea spectrum ($B-R=1.57$) for correcting
the de Vaucouleurs fits, and a Hubble-type Sab ($B-R=1.34$) for the
exponential fits. A power law SED with $\alpha=-1$ was assumed for
the BL~Lac nuclei.

For the filter conversions, the redshifted SED was convolved with the
transmission curve for the F28$\times$50LP filter and then normalised
to the best-fit host galaxy magnitudes. K corrections were performed
assuming $H_0 = 70$~km~s$^{-1}$Mpc$^{-1}$, $\Omega_M = 0.3$,
$\Omega_{\lambda} = 0.7$ (the cosmology used throughout).

The level of uncertainty in the host galaxy spectral corrections is potentially 
very large. For sources at these redshifts, the F28$\times$50LP filter
samples well below the rest-frame 4000\AA~ break, and hence is highly
sensitive to the star formation history of the galaxy. Uncertainties
may be as large as 0.6 mag if the high-redshift host galaxies have 
colours closer to those of spiral galaxies, rather than the red
colours observed at low redshift \citep{Scarpa2}, or even as large as
1~mag if they contain a strong starburst component. The quoted absolute
magnitudes should be interpreted with this in mind.

Galactic extinction was corrected using the relation to HI column
density from \citet{Shull}: $E(B-V) = log N(HI)/(21.72~cm^{-2}mag^{-1})$

\section{Host Galaxy Results}\label{properties}

The results of the fits for the single-component galaxy models and the PSF-only
model are given in Table~\ref{twocomp}. Table~\ref{threecomp} shows
the results for the composite galaxy models. 

\subsection{Host Galaxy Brightnesses}

The average absolute R-band magnitude of the five resolved host galaxies ---
0820+225, 1144--379, 1249+174, 1422+580, and 1533+535 --- for the de~Vaucouleurs model fit,  
K corrected assuming an early-type spectrum, is $M_R=-24.3$~mag. 
This is much brighter than the average magnitude of $M_R=-22.85$~mag
measured for the lower redshift BL~Lac host galaxies in \citet{Urry}.
This strongly suggests that there has been significant luminosity
evolution between the two samples, even taking into account the
presence of the (high) upper limits in this sample. The average
magnitude remains high even if a much bluer spectrum is used for
K corrections: $-23.7$~mag for a Hubble-type Sab spectrum, and
$-23.51$~mag for a 1~Gyr-old burst population.

\subsection{Host Galaxy Morphologies}

We were able to resolve complex morphologies in two cases. Both 0820+225 and  
1422+580 clearly have bulge-dominated host galaxies. 
The effective radii derived from almost all of the de~Vaucouleurs fits are
very large, although the sizable errors on these radii encompass more
sensible, smaller values. Nonetheless, the derived radii of
0820+225 and 1533+535 are unphysically large. These large
scale sizes may indicate that some of the host galaxies in this sample may
have substantial disk components as well as bulges. The bulge~+~disk
host galaxy model is preferred with high significance for 0820+225, and with
marginal significance for 1422+580 and 1533+535. 

Two of the unresolved sources, 0235+164 and 1308+326,
show a significant amount of extended light which is fit well by an
exponential disk component. This light can also be accounted for as
resulting from an imperfectly subtracted sky, at the limit of the errors
in the sky subtraction (and, in the case of 0235+164, at the limit of
the potential variation in the PSF). As such, the two are considered
non-detections. However, many of the host galaxies show the same excess of
extended light above the best-fit de Vaucouleurs profile. It is
possible that these `non-detections' are actually disky hosts on large scales.
These sources would require longer integration times to confidently
resolve the host galaxies. 

\subsection{Close Environments and Evidence of Interactions}
\label{environ}

Most of the BL~Lacs in this sample appear to have close companions, 
nearby irregular structure, or both. 1533+535 is the only source that
has no such observable features. It is impossible to judge in the case
of 0235+164
because of the known intervening system. Discounting this BL~Lac, 
five out of the remaining eight have apparent companions within a
projected distance of 50~kpc, and four of the eight show signs of
recent interactions, either companions within 10~kpc or irregular structure.

This sample differs from the broader BL~Lac sample of
\citet{Urry}, which exhibited close companions within 50~kpc
in 47\% of cases, rather than the 62.5\% as observed here. Very few of
the broader sample showed signs of recent interactions, compared to
50\% of this sample.

\subsection{Notes on Individual Objects}

{\it 0138--097}
The host galaxy of this BL~Lac appeared to be marginally resolved 
using the composite stellar PSF. With two of the alternate stellar
PSFs, however, it produced fits which were not sufficiently better
than the PSF-only fit. Thus, it is considered unresolved. 
The difficulty in resolving this host galaxy is compounded by the
presence of a bright companion object 1$^{\prime\prime}\!\!$.44
(10.5~kpc) to the southwest of the nucleus, and more structure within
$5\arcsec$. 

{\it 0235+164:}
The field of this object is complex, with bright structure extending
to $\sim3\arcsec$ ($\sim30$~kpc) from the nucleus. This structure 
has been well-studied in the past (e.g. Stickel {\it
et al.} 1988; Yanni, York \& Gallagher 1989; Abraham {\it et al.}
1993; Falomo 1996), and is known to be an intervening
system with redshift $z=0.524$.
These STIS data resolve the complex structure in significantly more
detail than previous observations, but we leave its analysis to a
later work.

Fortunately the extended structure is concentrated to the south of the
nucleus, leaving most of northern half clear for our host
galaxy analysis. A host galaxy appeared resolved using the composite
stellar PSF, however when an alternate stellar PSF was used and the
background was over-subtracted to its
uncertainty limit, it was not preferred over the PSF-only
fit with sufficient confidence, and so is considered unresolved.

{\it 0820+225:}
The host galaxy of this source is the best resolved of the sample
thanks to the relatively faint nucleus. Although the host has a strong
bulge component, a model including both bulge and disk components
provides a significantly better fit than the bulge-only model if effective radius of the
bulge is restricted to $\le 20$~kpc.
A compact companion galaxy is also resolved at
$\sim$0$^{\prime\prime}\!\!$.7 (5.5~kpc) to the west of the nucleus. 

{\it 1144--379:}
This host galaxy is resolved, although not sufficiently well to
distinguish between a bulge- or disk-dominated morphology.
A very faint, arc-like wisp is resolved at $\sim$2$^{\prime\prime}\!\!$.25
(18~kpc) to the northwest of the nucleus. 

{\it 1249+174:}
This host galaxy is definitely resolved; however, there was
insufficient information to constrain its morphology.
A bright, elongated structure is resolved extending from 
$\sim$0$^{\prime\prime}\!\!$.8 to $\sim$1$^{\prime\prime}\!\!$.8
($\sim$5 to 13~kpc) to the northwest of the nucleus, while two compact
objects are visible at $\sim10\arcsec$ from the BL~Lac.

{\it 1308+326:}
The host galaxy of this source is well fitted by an exponential disk
profile. However, it cannot be confidently distinguished from the PSF
if the background is over-subtracted to its
uncertainty limit, and so it is considered unresolved.
Two bright compact companion objects are within $6\arcsec$ (48~kpc) of
the BL~Lac, while a number of fainter objects are within half that distance.

{\it 1422+580:}
This host galaxy is well-resolved, and a de~Vaucouleurs profile
provides a significantly better fit than an exponential disk
profile. A host galaxy model including both bulge and disk components 
provides a marginally better fit. The host galaxy appears smooth and
is clearly elongated, with an axial ratio of 0.8.
Several compact companion objects are in the near vicinity of the BL
Lac, some faint ones within $\sim1^{\prime\prime}\!\!$.5 (13~kpc).

{\it 1533+535:}
Another well-resolved, smooth-looking host galaxy, although in this case the
morphology could not be constrained. This may be due to the presence
of a disk component, as the best-fit de~Vaucouleurs effective radius
is unphysically large. A de~Vaucouleurs~+~exponential profile gives
a marginally better fit than the de~Vaucouleurs-only model.

{\it 2240--260:}
The host galaxy of this source is unresolved. A few compact objects
lie to the northeast, $\sim6\arcsec$ (45~kpc) from the BL~Lac.

\section{Host Galaxy Evolution}\label{evolution}

To study the evolution of BL~Lac host galaxies we combine the STIS
observations described above (hereafter referred to as the high-z
sample) with the lower-redshift BL~Lacs observed in the WFPC2 snapshot
survey (those with $z \le 0.6$; hereafter the low-z sample). The
results of the host galaxy analysis for these
63 low-z BL~Lacs are listed in Table~\ref{snaplist}, and the details
of the analysis described in \citet{Scarpa}. 

The STIS F28$\times$50LP filter samples the rest-frame U to V bands at
redshifts $z > 0.6$, and so is particularly sensitive to star
formation. This fact has enabled us to detect BL~Lac host galaxies at
greater distances than previously achieved.
However, the star formation histories of the high-z BL~Lacs are unknown, 
and thus the filter conversions and K corrections are highly uncertain.
Any measurement of luminosity evolution in absolute magnitude space
would be subject to large errors.
However, from NICMOS imaging of 10 low redshift BL~Lacs 
\citep{Scarpa2}, we know that these host galaxies have 
colours consistent with old stellar populations. At the same time,
conversion from F702W to F28$\times$50LP magnitudes at lower redshifts
is far less sensitive to star formation history than conversion from
F28$\times$50LP to F702W at higher redshifts.
To minimise correction uncertainties we perform the analysis of host
galaxy luminosity 
evolution using the measured F28$\times$50LP apparent magnitudes of the
high-redshift sample, corrected only for galactic extinction, and 
the derived F28$\times$50LP apparent magnitudes of the low-redshift
sources, converted from the measured F702W apparent magnitudes.
The filter conversions used for the low-z sample are not fixed, but rather
varied within the fitting process (described in \S\ref{evolcen}).

The host
galaxy brightnesses of the combined sample are then compared to models
of galaxy luminosity evolution. These models are derived using
the GISSEL evolutionary synthesis models of \citet{Bruzual} with a
Salpeter IMF \citep{Salpeter}, with the 
{\em Padova 2000} library of isochrones \citep{Girardi},
and the BaSeL standard stellar library \citep{Lejeune}.

Figure~\ref{hubbleevol} shows the STIS F28$\times$50LP Hubble diagram for the
host galaxies, with the low-z sample filter conversions performed
assuming a solar metallicity burst population calibrated to match
the typical early-type colour of $B-V=0.96$ \citep{Fukugita}. 
Also shown is the line for a non-evolving population with the same
early-type spectrum ({\it dashed line}), and a non-evolving late-type
spectrum (solar metallicity, $age=8$~Gyrs, star formation rate decay exponent $\tau=-3$)
matching the Hubble type Sab colours of $B-V=0.57$ 
(Fukugita, Shimasaku \& Ichikawa 1995; {\it dotted line}). 
The high redshift host galaxies appear to be more luminous than
predicted by the non-evolving models. 
However, the large error bars and the presence of non-detections in
both the high-z and low-z samples necessitate rigorous testing. 

\subsection{$\chi^2$ Fitting with Censored Data}\label{evolcen}

The presence of host galaxy non-detections in some of the higher
redshift objects means that the apparent luminosity evolution may just
be a selection effect. Great care must be taken in dealing with the
non-detections, and for this we employ survival analysis techniques. 

The technique developed to apply survival analysis to the fitting of
evolution models is derived from the EM algorithm \citep{Dempster}.
This algorithm uses maximum
likelihood to estimate the regression coefficients of a 2-dimensional
data set, and assumes that the censored data points have a normal
distribution about the regression line. This regression line is
determined iteratively, with the estimates for the weighting of the 
censored data points updated at each iteration.
The assumption of a normal distribution 
makes the EM method less general than other, non-parametric methods, 
but given that the host galaxy luminosities have a roughly-normal 
distribution at low redshifts, this assumption seems reasonable.

As we are fitting the apparent magnitudes rather than the absolute
magnitudes, we can't perform a simple regression analysis.
Instead of iterating over regression line estimates, we iterate over
evolution models. If we assume that each galaxy consists of a single,
co-evolving population, then each model is defined by two parameters:
magnitude zero point and either population age or formation redshift
($z_{form}$) for the non-evolving and passively evolving models respectively. 

Each iteration in population age or formation redshift implies a set
of filter corrections for the low-z sample. Although there are only
small differences in the filter correction between different galaxy
models at low redshift, we wish to calculate $\chi^2$ given the
measured F702W magnitudes, not the inferred F28$\times$50LP
magnitudes, for each evolution model. With this in mind, we apply a
new, self-consistent spectral correction for each new population age
or formation redshift, rather than assume the same correction for all
iterations. For the moment, we assume solar metallicity in all models. 

\subsubsection{Single-Population Passive and Non-Evolving Models}

The passively evolving stellar population yielding the best fit
to the F28$\times$50LP apparent magnitudes of our sample has a
formation redshift of $z_{form} = 1.8$, with a 
reduced $\chi^2$ of 1.37. The $1\sigma$ uncertainties are $+0.7/-0.3$
and the $3\sigma$ uncertainties are $+1.0/-0.5$.
A better fit by a model with high
formation redshift can be ruled out with high certainty --- $z_{form}=3.0$
yields a reduced $\chi^2$ of 1.98, while $z_{form}=5$ gives a
reduced $\chi^2$ of 2.35. An F-test (72 data points and 2 degrees
of freedom) indicate a worse fit than the $z_{form}=1.8$ model with
95\% and 99\% significance respectively. 

The data are also reasonably well fit by an extremely young non-evolving
population. The best fit is for a population of age 0.02~Gyr, with 
reduced $\chi^2 = 1.51$ --- a
worse fit than the passive model, although without statistical
significance. A population as young as 1~Gyr provides a
poor fit to the data, with reduced~$\chi^2 = 2.13$: 
worse than the best passive model at the 97.5\% significance level,
and worse than the 0.02~Gyr model at the 95\% significance level. 
A population of age 2.5~Gyrs gives a reduced~$\chi^2$ of 2.38; worse
than the passive model at the 99\% level. Similar results are obtained
for models of non-evolving late-type galaxy spectra with the same
$B-V$ colours as the burst models. 

The young populations needed to fit the data 
do not agree with the colours observed in BL~Lac host galaxies at low
redshift. From \citet{Scarpa2}, we know that these
galaxies have $R-H = 2.3 \pm 0.3$, consistent with an old stellar
population. At the blue-most limit, these colours imply $age \sim
5$~Gyrs for a burst population. 
Fitting the normalization of a non-evolving model with $age =
5$~Gyrs results in a reduced~$\chi^2$ of 3.05. An F-test indicates that 
the passive evolution model gives a better fit than the 5~Gyr 
non-evolving model at the 99.9\% significance level.

Figure~\ref{hubblemodels} shows the Hubble
diagrams with the best-fit passive and non-evolving models
({\it top} and {\it middle}). In each case, the low-z host galaxies are converted from
F702W to F28$\times$50LP magnitudes assuming the corrections implied
by the evolution model plotted with a {\it solid line}.
Visual inspection of the Hubble diagrams shows that the non-evolving
model predicts galaxies about one magnitude too faint at $z > 0.6$. Even
much younger non-evolving models (1~Gyrs; {\it middle, dashed line},
fail to reproduce the high redshift colours,
although an unphysically young model (0.02~Gyrs; {\it middle, dotted line}),
does fit.
Passive evolution models with high formation redshifts similarly
underestimate the brightness of the $z > 0.6$ host galaxies, as can be
seen by the $z_{form}=5$ model ({\it top, dashed line}).

\subsubsection{Dual-Population Evolution Model}

While the above results indicate that the luminosity evolution is
better fit by a model that includes a rapidly evolving stellar
population, it
may not be the entire population that is evolving. If an old
stellar population is already in place, then a fresh burst of star
formation involving only a fraction of the galaxy's mass may also
reproduce the observed evolution.
We fit the data with a model that comprised a passively evolving
stellar population superimposed with a second population. This second
population remains at constant (low) age down to a given redshift,
after which it also evolves 
passively, to simulate a normally evolving galaxy with some
degree of ongoing star formation, which dies out at some point.
The model now includes five parameters: magnitude offset, formation
redshift of the primary population ($z_{form}$), the point at which the
secondary population begins to evolve passively ($z_{*}$), 
percentage mass of the secondary population relative to the primary,
$m_{*}$, and the initial age of the secondary population,
$age_{*}$. 

A wide range of parameters provided fits of similar quality to the
best-fit one-component model. For a primary population with a high
formation redshift ($z_{form} \gtrsim 4$), a secondary population of
at least 1\% of the mass of the primary (depending on
$z_{*}$ and $age_{*}$) produces reasonable fits. For a small secondary component
of 1\% to 3\%, $z_{*}$ can be as low as $\sim 0.5$ while still 
agreeing with the red colours observed in low-redshift BL~Lac host
galaxies. The best fit was found for $m_{*}$ of 2\%, $z_{*}=0.6$,
$age_{*}=1$~Gyr, and $z_{form} = 5$, with reduced~$\chi^2 = 1.29$. 

The bottom panel of figure~\ref{hubblemodels} shows the best-fit dual-population
evolution model ({\it solid line}). The {\it dashed line} in this plot
shows the evolution of the $z_{form}=5$ base population.

Although we can't distinguish between single- and dual-population
models, we can say with high confidence that the observed luminosity evolution is
significantly better represented by models which include star
formation activity in the range $0.5 < z \lesssim 2.5$, which diminishes
rapidly or ceases altogether below $z \sim 0.5$, than it is by models in
which the galaxies are undergoing steady star formation to low
redshift, or by models in which all star formation activity had ceased
by $z \sim 2.5$.

\subsection{Effects of Metallicity and Dust}

Given our knowledge of the red colours of low redshift BL~Lac host
galaxies, these data show conclusively that higher redshift BL~Lac
hosts are both significantly bluer and significantly brighter than
their low redshift counterparts. This change is well modelled
as evolution of the stellar population, however evolution in
metallicity or the presence of dust may also play a part.

\subsubsection{Evolution in Metallicity}

The evolution results given so far are for populations with constant, solar
metallicities. If the galaxies at higher redshift have much 
lower metallicities, the data can be fitted with 
older stellar populations. We simulate non-solar metallicities  
using the models of \citet{Westera}. 

Taking the host galaxies with $z > 0.6$ to have extremely low
metallicity --- 0.02~solar --- and those with $z < 0.6$ to have 
solar metallicity, the best-fit passively evolving population has
$z_{form}=2.3~^{+0.8}_{-0.7}$ (1$\sigma errors$), with reduced~$\chi^2=1.28$, better 
at 97\% confidence than a $z_{form}=5$ model (reduced~$\chi^2=2.07$).
The same metallicity evolution allows non-evolving populations to 
fit better. A population with $age=1$~Gyr gives reduced~$\chi^2=1.99$, 
worse than the best-fit passive model with only 95\% confidence 
(instead of 97.5\% for a constant, solar metallicity). 
However this population still
does not satisfy the low-redshift $R-H$ colour constraint.

Of course, a sudden transition from 0.02~solar to 1~solar metallicity
at $z=0.6$ is not realistic; however, this scenario provides a limit on
the extent to which metallicity evolution can affect the results.

\subsubsection{Effects of Dust Reddening on the Fits}

The presence of large amounts of dust in the host galaxies could
potentially result in extreme reddening --- 
enough to give a young population the
$R-H$ colours observed in low redshift BL~Lacs. However, there is a
limit to how much dust can be added, as 
the preferential extinction caused by dust at shorter wavelengths 
makes the high-redshift population fainter, worsening the fit. 

We applied models of dust extinction from \citet{Ferrara} to our 
spectral evolution models to determine whether a dust-reddened, 
non-evolving population could produce a
reasonable fit to the data, with the constraint that the low-redshift
($z < 0.4$) host galaxies have $R-H = 2.0$ (the bluest possible based
on the NICMOS observations), after dust reddening. 
The extinction model which achieved this most efficiently was one
based on the Milky Way extinction curve, assuming an elliptical galaxy
geometry with a Jaffe density profile, and a central V-band optical depth of
$\tau_V(0) = 2$. With this model, a non-evolving, $1~Gyr$-old population 
has $R-H=2$ at low redshift, and fits our Hubble diagram with
reduced~$\chi^2=2.21$. This is worse than
the best-fit passive model at the 98\% level by the F-test. To
substantially improve this fit (worse only at the 95\% level), 
a 0.6~Gyr population is needed, which, with this dust model, has $R-H = 1.7$ 
--- certainly far too blue for the low-redshift host galaxies.

Thus, no (non-evolving) dust
model can make a non-evolving stellar population fit the data. 
The level of dust reddening needed even for these relatively
poor fits, $\tau_V(0) = 2$, is at the upper limit of dust optical
depths estimated in ellipticals by \citet{Wise} or face-on spirals by \citet{Kuchinski},
and is substantially higher that the limit predicted for ellipticals by
\citet{Goudfrooij} of $\tau_V(0) < 0.7$, who find more typical values
of $\tau_V(0)\sim 0.2-0.3$.

\subsubsection{Evolution in Dust Content}

It is possible to reproduce the observed luminosity evolution for a 
non- or weakly-evolving population if dust content is allowed to 
increase rapidly between $z\sim 0.8$ and $\sim0.4$. Large changes 
at lower redshifts fail to fit the weak evolution seen at low-redshift, 
and extending the dust evolution to higher redshifts means 
that there is too much attenuation to 
reproduce the bright host galaxies at $z\gtrsim0.6$.

Assuming that the galaxies are relatively dust-free at $z\ge 0.8$, and
that optical depth in dust increases linearly to $z=0.4$,
the minimum final optical depth required to produce a reasonable
fit to the data and to satisfy the low-z $R-H=2$ constraint is 
$\tau_V(0) = 8$. This allows a non-evolving
model as young as 0.8~Gyrs to fit the data with reduced~$\chi^2=1.54$,
within 1$\sigma$ of the best-fit passive evolution model.

Of course, $\tau_V(0) = 8$ is extremely high ---
over an order of magnitude higher than the typical values determined  
for local ellipticals, and several times higher than those determined
for star-forming galaxies (e.g. Calzetti 2001).
If we assume a local optical depth of $\tau_V(0) = 2$, the upper limit 
predicted for ellipticals, then the best-fit non-evolving evolution
model gives reduced~$\chi^2 = 2.01$ --- still worse than the
best passive model at the $\sim$95\% level.

Approximately same level of dust evolution is required to make a
passively evolving population with high formation redshift 
fit the data. A population with $z_{form}=5$ requires a local optical
depth of $\tau_V(0) = 10$ for reduced~$\chi^2 = 1.52$, while the same
population with $\tau_V(0) = 2$ at $z=0$ gives reduced~$\chi^2=2.09$ 
--- worse than the $z_{form}=1.8$ model at 97\% confidence.

A concurrent evolution in metallicity will allow these models to fit
the data with a smaller increase in dust content. Using the metallicity 
evolution described above, the minimum local optical depth needed for 
a 2~Gyr-old non-evolving or a $z_{form}=5$ passively evolving population 
to fit the data (within 1$\sigma$ of the best passive model) is $\tau_V(0)= 5$ 
(reduced~$\chi^2 \sim 1.5$). This is still an extreme increase in dust
content. If $\tau_V(0)=2$ locally, then these models are
worse than the best passive model at the $\sim$95\% confidence level.

Figure~\ref{dustmetalevol} shows the Hubble
diagram with both the non-evolving and passive evolution models,
dust and metallicity evolution included. 

\subsection{Absolute Luminosity Evolution}\label{lumevol}

The lack of extended emission line light or a big blue bump has
made it possible to use the extremely sensitive F28$\times$50LP filter 
in this study. However, to compare our evolution results directly to
other studies, it is necessary to determine this evolution in absolute
near-infrared magnitudes.
This result depends strongly on the K corrections used. To find the
full range of possible magnitude-redshift gradients, we perform
regression analysis on absolute K-band magnitudes using the range of 
K corrections implied by the full range of possible
evolutionary models from the above analysis. 

We use the EM algorithm to perform regression analysis while accounting for
the censored data points, via the IRAF task EMMETHOD. This task by
itself does not take error bars into account. As the uncertainties in
host galaxy magnitudes increase with redshift, they are an important
consideration. To account for them, we use a Monte Carlo
approach. One hundred simulated data sets were generated, randomizing each
detected data point over a composite normal distribution, with the
widths of the upper and lower halves of the distribution defined by
the upper and lower error values for that data point.
The regression coefficients and their uncertainties were
determined for each of these 100 simulated data set with EM method, and the
final coefficients taken to be the mean values. The uncertainty values
were taken to be the mean of the individual uncertainties in the
regression coefficients, added in quadrature with the standard
deviation of the coefficients across the 100 simulations.

The steepest slope occurs when K corrections are derived from the 
evolutionary model with the oldest possible population at high
redshift (at the 3$\sigma$ level, assuming no dust or metallicity
evolution) --- the passive evolution model 
with $z_{form} = 2.8$. The shallowest slope is in the case of the
youngest possible high-z population (also at the 3$\sigma$ level) ---
a passive model with $z_{form} = 1.3$. Unsurprisingly, the best-fit
passive evolution model, with $z_{form} = 1.8$, gives a gradient about
halfway between these. The results of the regression analysis using
these three sets of K corrections are shown in Table~\ref{regresstable}.

A similar range of gradients are obtained if we assume a secondary burst
component added to an older population, with a lower limit of $0.59$~mag/z 
for a passive, old population plus a 0.1~Gyr burst involving 3\% of the galaxy's mass
which itself passively evolves below $z=0.5$.
For a constant-age, old population,
or a population with high formation redshift we obtain $\sim1.7$~mag/z.
To obtain high
redshift K-band magnitudes consistent with no K-band luminosity
evolution, the stellar populations at $z > 0.6$ must have ages less
than 0.2~Gyrs. Such star formation histories can only be made to
fit the observed Hubble diagram and satisfy the low-redshift colour
constraints with extreme levels of dust evolution 
(from $\tau_V(0)\sim0$ to $\tau_V(0)\sim5$ between $z=0.8$ and 0.4).

Using the best fit to the Hubble diagram to derive K corrections, and adding
in quadrature the error in the regression analysis to the probable
range of slopes from different K corrections, we determine the overall
K-band luminosity evolution of BL~Lac host galaxies to be 
$1.0 \pm 0.45$~mag/z.
Figure~\ref{mzregress} shows 
absolute K-band magnitude versus redshift, K corrected assuming a
passively evolving population with $z_{form}=1.8$.
Also plotted are the best-fit linear regression lines representing
the magnitude-redshift gradients
derived assuming the three different passive evolution models.

\section{Discussion}\label{discussion}

We detect luminosity evolution in BL~Lac host galaxies, consistent 
with the primary stellar population evolving strongly from 
$1.5\lesssim z \lesssim 2.5$, or with a smaller 
burst of star formation evolving 
from $z>0.5$, and inconsistent with a non-evolving population or a 
population evolving passively from high redshift. Interestingly, this formation 
epoch for the primary population corresponds roughly to 
the epoch of maximum quasar activity (e.g. Maloney \& Petrosian 1999;
Shaver {\it et al.} 1999). This is consistent with a scenario in 
which black hole feeding and growth --- i.e., the epoch of maximum 
AGN activity --- coincides with the formation of 
the surrounding bulge \citep{KormendyG}.

Radio galaxies \citep{Djorgovski, Spinrad, Spinrad2, Lilly4, Lilly2, 
Aragon-Salamanca, McLure1} and the hosts of radio-loud quasars 
\citep{Lehnert, Kukula, Hutchings3} exhibit similar evolution, 
probably dimming by of order 1 magnitude in K between $z=1$ and 0.
Although this evolution is not as well constrained, 
it is consistent with that seen in our sample. This is in agreement with 
the unified model of AGN in which radio galaxies, radio quasars and 
blazars are expected to have the same host galaxy properties.
In contrast, BCGs \citep{Aragon-Salamanca} other quiescent early-types 
\citep{Lilly1, Stanford, Bell} and 
radio-quiet quasar hosts \citep{Kukula, Rix, Ridgway2} exhibit flat, or 
even negative luminosity evolution, with local galaxies appearing as bright or 
brighter than those at $z\gtrsim 1$. Thus, the hosting of a powerful radio jet seems 
to be linked to a galaxy's luminosity evolution.

The luminosity evolution of BL~Lac host galaxies is 
unambiguous, not affected by the biases inherent in 
other radio-loud classes (see \S\ref{intro}).
We can therefore ascribe the evolution observed in BL~Lac hosts
to a real change in galaxy luminosities, which in turn grants us more
confidence that the 
evolution observed in other radio-loud hosts is real.

Furthermore, BL~Lacs allow us significantly more confidence that the
observed evolution is a luminosity evolution in individual host galaxies.
There is mounting evidence to suggest that BL~Lac
objects are long-lived, possibly with lifespans on the same timescales as 
galaxy evolution. The most compelling is the flat, possibly negative number
density evolution of many BL~Lacs \citep{Morris, Giommi, Bade, Perlman2}. 
There are as many, if not more BL~Lac objects now than during the
epoch of peak quasar activity. 
The low accretion rates in BL~Lacs (less than 1\% Eddington; O'Dowd, Urry \&
Scarpa 2002; Wang, Staubert \& Ho 2002) also suggest that
BL~Lacs may burn stably for extended periods. If the jets of BL~Lacs obtain most 
of their power from the rotational energy of the black hole 
(e.g. Blandford \& Zjanek 1977, Meier 2002), they are expected to have lifespans of the order of gigayears
\citep{CavaliereM}. 
Long-lived emission is also supported by their dynamically evolved state at low redshift
(e.g., Urry {\it et al.} 2000)
compared to the high frequency of interaction remnants and close
companions at higher redshifts (see \S\ref{environ}).
Thus, with BL~Lacs we may be observing luminosity evolution intrinsic to the host 
galaxies themselves, rather than evolution in the processes which select galaxies as hosts of 
radio-loud AGN.

Our results are consistent with and confirm the results of the 
ground-based study of \citet{Heidt}, who find evidence 
of luminosity evolution in BL~Lac host galaxies, although do 
not constrain this evolution. 
Of the sources which overlap the two samples 
--- 0820+225, 1249+174, 1422+580 and 2240--260 --- 
Heidt {\it et al.} detect the host of 0820+225. Our host galaxy 
magnitudes for this source agree within error, although our 
effective radii do not. 
We believed that this stems from the uncertainty and 
width of the ground-based PSF (of order that of the host itself 
in some cases). 

\subsection{Possible Causes of Luminosity Evolution}
\label{causes}

Both quiescent, early-type galaxies and radio-quiet AGN host galaxies 
are gaining luminous mass up to the present, in line with the predictions
of hierarchical clustering models \citep{Kauffmann}, therefore the
luminosity of these galaxies does not change with time. This does not
appear to be the case with the host galaxies of radio-loud AGN, whose
dimming over cosmic time is apparently not offset by
the accretion of significantly amounts of new material.  
A number of effects may contribute to this.

\noindent {\bf Evolution linked to emission processes:}
Low redshift radio-loud AGN used in evolution studies are dominated by 
sources with low accretion onto their SMBHs, such as the FR~I and
low-power FR~II radio galaxies in the 3C sample (in which luminosity
evolution has been best studied). These are also the proposed parent
population of BL~Lacs \citep{UrryP}. If a large 
fraction of these AGN are long-lived, jet-dominated sources, then we 
select against active recent merger histories when we observe them 
locally. Continued merger events following an AGN's formation epoch 
will refuel the nucleus after depletion of the initial supply of gas
and possibly disrupt stable, jet-dominated emission \citep{Cavaliere}, 
as well as add luminous mass to the galaxy, offsetting the dimming due 
to an aging primary stellar population.

\noindent {\bf Evolution Linked to Black Hole Properties:}
While the production of a radio jet is likely to be linked to the 
properties of the central SMBH, it is uncertain
exactly which black hole properties are important in determining radio
loudness. There is some evidence that high black hole mass plays a part
\citep{Franceschini2, Lacy, Nagar, Jarvis, ODowd, Dunlop}, although 
this may just be a selection effect \citep{Woo, Oshlack, Ho}.
Black hole spin seems the more likely candidate as the critical 
determinant \citep{Meier, ODowd2}. On the other hand, high host galaxy
mass is expected to dampen radio emission \citep{Bicknell}.
Galaxies that have extreme SMBH properties relative to their mass 
should favor the production of radio jets. Such SMBHs may be expected 
to preferentially occur in those galaxies whose 
formation was completed by higher redshifts.

In the case of SMBH mass, hierarchical models 
\citep{Kauffmann} predict that earlier-formed
galaxies will have higher SMBH-to-host galaxy mass ratios, 
due to the larger amounts of cold gas available for accretion at
earlier epochs. A correlation has even been observed between the
fraction of a galaxy's mass in the black hole and the age of its
stellar population \citep{Merrifield}. 

In the case of spin, a black hole of a given mass that is
formed from a small number of major merger events is likely
to have a higher spin than one of the same mass formed 
from many less massive mergers, as the angular momenta of these 
lesser interactions will tend to cancel out unless they are 
correlated \citep{Cavaliere}. Locally, this means black holes of 
given mass which formed during the epoch of major mergers, at 
$z\sim$1 -- 2, will have higher spin than whose whose formation 
proceeded to the present.

According to the scenarios outlined above, when we select radio
sources at low redshift, we are dominated by low-accretion AGN with
extreme SMBH properties that are preferentially produced at higher redshifts,
and that should have had relatively quiescent merger histories since
the epoch of formation of their central engines. We have more
credible evidence that radio-loud AGN do exhibit just such merger
histories, supporting the idea that the properties of radio-loud AGN
SMBHs are typically defined at high redshift.   

\subsection{The Merger History of Radio Hosts}

In \S\ref{environ} it was found that many of the BL~Lacs in the
high-z sample have either close companions (62.5\%) or signs of recent
interactions (50\%). Fewer of the broader sample of \citet{Urry}
had close companions (47\%) and only a handful ($<$10\%) showed signs of
interactions. 
This supports the idea that the high-z BL~Lacs are closer to
their formation epoch, and that the observed
luminosity evolution is linked to the merger history of these host
galaxies. 

It is impossible to perform a such a comparison over a similar redshift 
range for other radio-loud AGN, as flux limits induce a power-redshift 
correlation. Locally, however, FR~I radio galaxies, which are 
thought to be the primary parent population of BL~Lac objects, 
do show fewer signs of recent interactions 
than the more powerful FR~IIs \citep{Heckman2, Smith1}, with 
the latter hosted by bluer galaxies and showing a much higher incidence of
distorted morphologies.

These observations, along with the scenarios outlined in \S\ref{causes}, point 
to a scheme in which radio-loud AGN are formed in major mergers, typically 
at high redshift, while those observed locally have evolved from a significantly
earlier formation epoch relatively free of further interactions.

Given this picture, it is expected that cluster environment will play a 
critical role in governing when and where radio-loud AGN form and survive.
Formation will be dictated by a balance between a high density 
environment (clusters for merger frequency) and low cluster velocity dispersion 
(high redshift clusters or local field, to permit full mergers; Makino
\& Ebisuzaki 1996). Survival as jet-dominated sources to low redshift
will favour lower density environments, 
as we require fewer ongoing merger events. 
BL~Lacs do prefer relatively poor cluster environments at low redshift ($z<0.4$),
of Abel richness class $\sim$0 to 1 \citep{FalomoPT, Pesce, Fried,
  Wurtz, Smith3}. At higher redshifts, their environments are not 
well known, although at $z > 0.6$ at least some are found in rich clusters 
\citep{Fried, Wurtz}. This is consistent with the picture in which 
BL~Lacs are preferentially produced in cluster environments at high redshift 
(when cluster velocity dispersions were lower), but survive only in the 
sparcer regions of these clusters. Reports on the environmental
evolution of other  
classes of radio-loud AGN are conflicting, as are the environmental differences 
between radio-loud and radio-quiet AGN.

\section{Conclusions}\label{conclusions}

Nine BL~Lac objects with $z > 0.6$ were imaged with HST, using the
STIS CCD and the F28$\times$50LP filter. The host galaxies of five of
these were resolved with high confidence. For all but 0820+225, these 
are the first detections of host galaxies in these sources, and include 
the two highest redshift BL~Lac host galaxies detected to date. These galaxies 
are bright,with an average absolute magnitude of
$<$M$_R> = -24.3$~mag assuming an early-type spectrum and
$-23.7$~mag assuming a late-type spectrum.

The morphologies of two of the galaxies were well resolved. Both 1422+580
and 0820+225 are better fit by a de~Vaucouleurs profile than by an
exponential disk profile, indicating bulge dominated hosts, as is
typical for radio-loud AGN. However, the host galaxy of 0820+225 was
significantly better fit by a composite model including both bulge and
disk components than by a bulge component alone. 
Many, including some of those technically considered
non-detections, show evidence of having extended, disky components.
A large fraction of the sample show evidence of close companions and 
distorted morphologies --- far more than are observed at lower redshift.

Combining the derived host galaxy magnitudes with the low redshift host
galaxies from \citet{Urry}, we measure luminosity evolution in a sample 
of 72 HST-imaged BL~Lac host galaxies spanning the range $0\lesssim z\lesssim1$.
We find that the Hubble diagram is well fit by a model of a
stellar population evolving passively from $z = 1.8~^{+0.7}_{-0.3}$. 
Alternatively, an older population with a small fraction ($\lesssim$3\%) 
undergoing active star formation to $z\sim0.5$ also fits the data
well. A single population undergoing passive evolution
from a high formation redshift ($z_{form} \gtrsim 5$) 
can be ruled out with high confidence.
By constraining the colours of the low redshift host galaxies based on
NICMOS observations, models including no luminosity evolution can
also be ruled out with
high confidence. Non-evolving and high-z$_{form}$ passive 
evolution models are still ruled out when reasonable models of dust and 
metallicity evolution are included.

By K-correcting the host galaxy magnitudes based on the range of
acceptable fits to the Hubble diagram, we determined an 
average K-band luminosity evolution of $1.0 \pm 0.45$
magnitudes per unit redshift. 

The fact that flux selection of BL~Lacs is dominated by beaming angle
rather than intrinsic brightness, together with the low levels of extended
emission from ionized gas and scattered nucleur light in these sources, 
mean that they are relatively free of the
biases affecting quasar studies. This increases our confidence 
that the observed luminosity evolution is
real, and in turn in the
luminosity evolution observed in other classes of radio-loud AGN.
We also have reason to believe that the observed evolution is
intrinsic to the galaxies themselves, as evidence
suggests that BL~Lacs may have lifespans comparable to the timescale for galaxy
evolution. 

The luminosity evolution of BL~Lac host galaxies is steeper than that
observed in radio-quiet host galaxies, brightest cluster galaxies, and 
quiescent early-type galaxies,
and is consistent with that observed in other radio-loud hosts.
The absence of luminosity evolution in radio-quiet and quiescent
galaxies suggests that dimming due to an aging population is offset by
the accretion of new material, as is predicted by hierarchical
clustering models. This appears not to be the case in
the host galaxies of radio-loud sources, which evolve
strongly from some redshift greater than $z\sim0.5$, but also must have
experienced significant star formation activity in the range 
$0.5 < z < 2.5$. This is further supported by the fact that BL~Lacs
with $z > 0.6$ are much more frequently in interacting systems, and
have more close companions than those at low redshift, which appear
dynamically evolved. These observations lead to the following
conclusions:

First, it seems that the physical conditions that allow a galaxy to
generate a radio jet are preferentially produced in a particular cosmic epoch,
albeit a broad one: $0.5 < z < 2.5$. This may be linked to the 
amount of gas available for fueling, which is also believed to define
the quasar epoch. However, to explain the radio-loud---radio-quiet dichotomy, 
it must also be linked to the production of the right SMBH conditions. 

Second, radio-loud AGN hosts at low redshift accrete less new material
after the epoch in which they seemingly form their SMBHs than
do other early-type galaxies. A number of effects may result in the 
selection of inactive recent merger histories when we select radio-loud AGN 
at low redshift; however, all point to a scenario in which many radio-loud AGN 
observed at low redshift have survived from a much earlier formation epoch.
 
The evolution observed in the stellar populations of radio-loud AGN
suggests the black holes which power these radio sources are born in
gas-rich interactions at $z > 0.5$. At low redshift, where samples 
are dominated by low-accretion sources, these observations suggest  
that we are
observing the last embers of the radio quasar epoch --- old AGN with
massive, rapidly-rotating black holes which have experienced little 
merger activity since the formation of their central engines. 


\acknowledgments
Support for this work was provided by NASA through grant numbers
HST-GO-09121.01-A and HST-GO-09223.01-A
from the Space Telescope Science Institute, which is operated by 
AURA, Inc., under NASA contract NAS5-26555.

\clearpage

\begin{deluxetable}{cccccccc}
\tablecaption{\label{targlist}}
\tablewidth{0pt}
\tablehead{
\colhead{Name} & \colhead{z} & \colhead{Type\tablenotemark{a}}
&\colhead{$m_R$(nucl.)\tablenotemark{b}}
&\colhead{$M_R$(nucl.)\tablenotemark{c}}
&\colhead{$m_R$(host)\tablenotemark{d}}   
&\colhead{Exp. Time (s)}
&\colhead{Obs. Date)}}
\startdata
 0138--097& 0.733 & L &17.7~~&$-$25.57& $>$19.57  &2554.0 &24/07/2001\\
 0235+164 & 0.94~ & L &18.90 &$-$25.03& $>$19.4~~ &5429.0 &18/07/2001\\
 0820+225 & 0.951 & L &20.0~~&$-$23.97& $>$21.25  &5473.0 &17/09/2001\\
 1144--379& 1.048 & L &18.0~~&$-$26.22& $>$22.72  &5615.0 &04/08/2001\\
 1249+174 & 0.644 & H &18.5~~&$-$24.43& $>$21.3~~ &2552.0 &21/07/2001\\
 1308+326 & 0.997 & L &18.1~~&$-$25.99& $>$20.13  &5551.0 &22/07/2001\\
 1422+580 & 0.683 & H &19.07 &$-$24.01& $>$21.7~~ &4644.0 &26/08/2001\\
 1533+535 & 0.89~ & H &18.60 &$-$25.19& $>$19.45  &4861.0 &14/09/2001\\
 2240--260& 0.774 & L &17.5~~&$-$25.91& $>$21.45  &2578.0 &27/07/2001\\
\enddata
\tablenotetext{a}{SED type: H=HBL ($F_{1keV}/F_{5GHz} >  5.5$) and
  L=LBL ($F_{1keV}/F_{5GHz} < 5.5$)}
\tablenotetext{b}{Apparent R-band magnitude of source, not extinction
  or K corrected.} 
\tablenotetext{c}{Absolute R-band magnitude of source, extinction
  corrected and K corrected assuming $\alpha=-1$, $H_0 =
  70$~km~s$^{-1}$Mpc$^{-1}$, $\Omega_M = 0.3$, $\Omega_{\lambda} = 0.7$.}
\tablenotetext{d}{Limit on apparent R-magnitude of the host galaxy from \citet{Urry}.}
\end{deluxetable}


\begin{deluxetable}{lllclclccc}
\tablecaption{Results of two-component host galaxy fits
\label{twocomp}
}
\tabletypesize{\tiny}
\rotate
\tablewidth{0pt}
\tablehead{
\colhead{Name}
&\colhead{Model\tablenotemark{a}}
&\colhead{m$_{LP}$~nucleus\tablenotemark{b}}
&\colhead{M$_R$ nucl.\tablenotemark{c}}
&\colhead{m$_{LP}$~host\tablenotemark{b}}
&\colhead{M$_R$~host\tablenotemark{c}}
&\colhead{r$_e$ (arcsec)\tablenotemark{d}}
&\colhead{r$_e$ (kpc)\tablenotemark{d}}
&\colhead{b/a\tablenotemark{e}}
&\colhead{$\chi^{2}$\tablenotemark{f}}}
\startdata
          &non-detect.         &~16.63~~$-$0.01/+0.01 &$-$26.78 &~$>$19.19                        &~$>-$25.31         &\nodata               &\nodata  &\nodata&0.971\\
0138--097 &(de Vauc.)          &(16.74)              &($-$26.67)&\phm{$>$}(20.01)                 &\phm{$>$}($-$24.33)&(4.2)                 &(30.6)   &\nodata&0.959\\
          &(expon.  )          &(16.69)              &($-$26.72)&\phm{$>$}(20.36)                 &\phm{$>$}($-$23.52)&(1.1)                 &(8.0)    &\nodata&0.962\\
\hline                                                                                                                                              
          &non-detect.         &~17.93~~$-$0.01/+0.02 &$-$26.14 &~$>$20.61                        &~$>-$25.04         &\nodata               &\nodata  &\nodata&1.055\\
0235+164  &(de Vauc.)          &(18.09)              &($-$25.98)&\phm{$>$}(20.80)                 &\phm{$>$}($-$24.69)&(2.4)                 &(18.9)   &\nodata&1.022\\
          &(expon.  )          &(17.97)              &($-$26.10)&\phm{$>$}(21.16)                 &\phm{$>$}($-$23.69)&(1.9)                 &(15.0)   &\nodata&1.018\\
\hline                                                                                                                                              
          &de Vauc.            &~20.38~~$-$0.13/+0.11 &$-$23.81 &~\phm{$>$}21.16~~$-$0.25/+0.40   &~\phm{$>$}$-$24.64 &~5.7~+14.2/$-$5.4     & 45.2    &1.0    &0.847\\
0820+225  &expon.              &~20.24~~$-$0.12/+0.16 &$-$23.95 &~\phm{$>$}21.61~~$-$0.80/+0.84   &~\phm{$>$}$-$23.54 &~1.0~\phn+0.3/$-$0.4  &  7.9    &1.0    &0.886\\
          &PSF only            &~20.28~~$-$0.05/+0.04 &$-$23.91 &\nodata                          &\nodata            &\nodata               &\nodata  &\nodata&0.928\\
\hline                                                                                                                                              
          &de Vauc.            &~18.00~~$-$0.12/+0.06 &$-$26.74 &~\phm{$>$}21.18~~$-$0.46/+0.38   &~\phm{$>$}$-$25.47 &~2.5~\phn+4.0/$-$0.9  & 20.2    &1.0    &1.521\\
1144--379 &expon.              &~17.99~~$-$0.09/+0.09 &$-$26.75 &~\phm{$>$}21.28~~$-$0.42/+0.20   &~\phm{$>$}$-$24.52 &~0.9~\phn+0.5/$-$0.3  &  7.3    &1.0    &1.547\\
          &PSF only            &~17.81~~$-$0.02/+0.03 &$-$26.93 &\nodata                          &\nodata            &\nodata               &\nodata  &\nodata&1.564\\
\hline                                                                                                                                              
          &de Vauc.            &~18.94~~$-$0.12/+0.08 &$-$24.09 &~\phm{$>$}21.22~~$-$0.25/+0.62   &~\phm{$>$}$-$22.71 &~1.1~\phn+2.3/$-$0.7  &  7.6    &1.0    &0.452\\
1249+174  &expon.              &~18.68~~$-$0.11/+0.08 &$-$24.35 &~\phm{$>$}22.23~~$-$0.24/+0.52   &~\phm{$>$}$-$21.29 &~0.8~\phn+1.7/$-$0.6  &  5.5    &1.0    &0.465\\
          &PSF only            &~18.72~~$-$0.04/+0.05 &$-$24.31 &\nodata                          &\nodata            &\nodata               &\nodata  &\nodata&0.488\\
\hline                                                                                                                                              
          &non-detect.         &~17.66~~$-$0.04/+0.04 &$-$26.47 &~$>$21.29                        &~$>-$24.59         &\nodata               &\nodata  &\nodata&0.776\\
1308+326  &                    &                      &         &                                 &                   &                      &         &       &\\
          &(expon.  )          &(17.69)              &($-$26.44)&\phm{$>$}(21.45)                 &\phm{$>$}($-$23.62)&(3.0)                 &(24.0)   &\nodata&0.762\\
\hline                                                                                                                                              
          &de Vauc.            &~18.56~~$-$0.08/+0.06 &$-$24.59 &~\phm{$>$}20.35~~$-$0.30/+0.22   &~\phm{$>$}$-$23.78 &~2.6~\phn+1.4/$-$1.1  & 18.4    &0.8\phn~$\pm$0.09~ &1.074\\
1422+580  &expon.              &~18.35~~$-$0.04/+0.06 &$-$24.80 &~\phm{$>$}20.88~~$-$0.10/+0.06   &~\phm{$>$}$-$22.81 &~0.9~\phn+1.1/$-$0.5  &  6.4    &0.75~$\pm$0.13  &1.140\\
          &PSF only            &~18.16~~$-$0.03/+0.03 &$-$24.99 &\nodata                          &\nodata            &\nodata               &\nodata  &\nodata&1.296\\
\hline                                                                                                                                              
          &de Vauc.            &~18.26~~$-$0.07/+0.06 &$-$25.59 &~\phm{$>$}20.63~~$-$0.24/+0.33   &~\phm{$>$}$-$24.66 &~3.4~\phn+3.0/$-$1.0  & 26.4    &1.0    &0.517\\
1533+535  &expon.              &~18.18~~$-$0.08/+0.08 &$-$25.67 &~\phm{$>$}21.13~~$-$0.50/+0.31   &~\phm{$>$}$-$23.61 &~0.8~\phn+0.6/$-$0.2  &  6.2    &1.0    &0.541\\
          &PSF only            &~17.99~~$-$0.02/+0.04 &$-$25.86 &\nodata                          &\nodata            &\nodata               &\nodata  &\nodata&0.624\\
\hline                                                                                                                    
          &                    &                      &         &                                 &                   &                      &         &       &\\
2240--260 &non-detect.         &16.64~~$-$0.02/+0.02  &$-$26.84 &~$>$20.58                        &~$>-$24.08         &\nodata               &\nodata  &\nodata&0.598\\
          &                    &                      &         &                                 &                   &                      &         &       &\\
\enddata
\tablenotetext{a}{Model fitted: PSF+de~Vaucouleurs, PSF+exponential
  disk, or PSF only; {\it non-detect} indicates fit with PSF alone where host galaxy
  is not resolved. Bracketed values indicate best host galaxy fits, although are
  technically non-detections.}
\tablenotetext{b}{Apparent magnitude of nucleus/host galaxy in STIS
  F28$\times$50LP band, not extinction corrected or K corrected.}
\tablenotetext{c}{Absolute magnitude of nucleus/host galaxy in Cousins
  R band, corrected for galactic extinction and K corrected assuming
  $\alpha=1$ power law for nuclei, early-type spectrum for de~Vaucouleurs hosts
  and late-type spectrum for exponential disk hosts, and
  $H_0 = 70$~km~s$^{-1}$Mpc$^{-1}$, $\Omega_M = 0.3$, $\Omega_{\lambda} = 0.7$.}
\tablenotetext{d}{Characteristic radius, assuming $q_0 = 0$ for kpc radii.}
\tablenotetext{e}{Axial ratio.}
\tablenotetext{f}{Reduced $\chi^2$ for fit, calculated comparing full two-dimensional model to the image.}
\end{deluxetable}


\begin{deluxetable}{ccccccccc}
\tablecaption{Results of three-component host galaxy fits
\label{threecomp}
}
\tablewidth{0pt}
\tabletypesize{\scriptsize}
\tablehead{
\colhead{Name}         &
\colhead{$M_R$(nucl.)\tablenotemark{a}} &
\colhead{$m_R$(nucl.)\tablenotemark{b}} & 
\colhead{Component\tablenotemark{c}} & 
\colhead{m$_{LP}$(host)\tablenotemark{d}} & 
\colhead{M$_R$(host)\tablenotemark{e}} & 
\colhead{r$_e$($\arcsec$)\tablenotemark{f}} & 
\colhead{r$_e$(kpc)\tablenotemark{g}} &
\colhead{$\chi^2$\tablenotemark{h}}}
\startdata
0820+225           &20.37    &$-$25.14   &de Vauc.&22.18    &$-$24.85    &1.4     &11.0      &0.811\\
                   &         &           &expon.  &21.63    &$-$24.75    &2.0     &15.8&\\
\hline                                                                                                                             
1144--379          &17.98    &$-$28.12   &de Vauc.&21.35    &$-$26.28    &1.1     & 8.9      &1.514\\
                   &         &           &expon.  &21.90    &$-$24.88    &1.3     &10.5&\\
\hline                                                                                                                             
1249+174           &18.71    &$-$25.54   &de Vauc.&21.91    &$-$23.37    &0.7     & 4.8      &0.446\\
                   &         &           &expon.  &22.79    &$-$22.08    &1.9     &13.1&\\
\hline                                                                                                                             
1422+580           &18.37    &$-$26.02   &de Vauc.&21.00    &$-$24.82    &1.4     & 9.9      &1.049\\
                   &         &           &expon.  &21.33    &$-$24.05    &2.4     &17.0&\\
\hline                                                                                                                             
1533+535           &18.25    &$-$26.90   &de Vauc.&21.10    &$-$25.52    &1.8     &14.0      &0.505\\
                   &         &           &expon.  &21.59    &$-$24.47    &3.0     &23.3&\\
\enddata
\tablenotetext{a} {Apparent magnitude of nucleus in STIS
  F28$\times$50LP band, not extinction or K corrected.} 
\tablenotetext{b} {Absolute R-band magnitude of nucleus, corrected for
  galactic extinction and K corrected assuming $\alpha=1$, 
  $H_0 = 70$~km~s$^{-1}$Mpc$^{-1}$, $\Omega_M = 0.3$, $\Omega_{\lambda} = 0.7$.} 
\tablenotetext{c} {de~Vauc. = de~Vaucouleurs (bulge) component;
  expon. = exponential disk component.} 
\tablenotetext{d} {Apparent magnitude of host galaxy in
  F28$\times$50LP band, not extinction or K corrected.} 
\tablenotetext{e} {Absolute R-band magnitude of host galaxy,
  extinction corrected and K corrected assuming early- and late-type
  spectra for de Vaucouleurs and exponential components respectively,
  with $H_0 = 70$~km~s$^{-1}$Mpc$^{-1}$, $\Omega_M = 0.3$, $\Omega_{\lambda} = 0.7$.} 
\tablenotetext{f} {Characteristic radius in arcseconds.}
\tablenotetext{g} {Characteristic radius in kiloparsecs, assuming $q_0 = 0$.} 
\tablenotetext{h} {Reduced $\chi^2$ for fit, calculated comparing full
  two-dimensional model to the image.} 
\end{deluxetable}


\begin{deluxetable}{lcccrclcc}
\tablecaption{The low-z sample: BL~Lac host galaxies from \citet{Urry}.
\label{snaplist}}
\tabletypesize{\scriptsize}
\tablewidth{0pt}
\tablehead{
\colhead{Name}
&\colhead{z}
&\colhead{Type\tablenotemark{a}}
&\colhead{m$_R$~nucleus\tablenotemark{b}}
&\colhead{m$_R$~host\tablenotemark{c}}
&\colhead{m$_R$~nucl.\tablenotemark{d}}
&\colhead{m$_R$~host\tablenotemark{e}}
&\colhead{r$_e$ (arcsec)\tablenotemark{f}}
&\colhead{r$_e$ (kpc)\tablenotemark{g}}}
\startdata
0118--272 & 0.559 & L & 15.78~~$\pm$~~0.10 & $>$19.09~~~~~~        &  $-$26.78 &  $>-$23.47 & ---               &  ~0.0~  \\
0122+090  & 0.339 & H & 21.98~~$\pm$~~0.25 &    18.88~~$\pm$~~0.04 &  $-$19.29 & ~~$-$22.85 & 1.05~~$\pm$~~0.10 &  ~5.08  \\
0158+001  & 0.229 & H & 18.38~~$\pm$~~0.06 &    18.27~~$\pm$~~0.03 &  $-$21.91 & ~~$-$22.29 & 1.90~~$\pm$~~0.10 &  ~6.96  \\
0229+200  & 0.139 & H & 18.58~~$\pm$~~0.35 &    15.85~~$\pm$~~0.01 &  $-$20.52 & ~~$-$23.41 & 3.25~~$\pm$~~0.07 &  ~7.97  \\
0257+342  & 0.247 & H & 19.18~~$\pm$~~0.30 &    17.93~~$\pm$~~0.01 &  $-$21.30 & ~~$-$22.85 & 1.75~~$\pm$~~0.12 &  ~6.78  \\
0317+183  & 0.190 & H & 18.28~~$\pm$~~0.05 &    17.59~~$\pm$~~0.01 &  $-$21.56 & ~~$-$22.46 & 3.25~~$\pm$~~0.10 &  10.31  \\
0331--362 & 0.308 & H & 19.03~~$\pm$~~0.10 &    17.81~~$\pm$~~0.02 &  $-$22.00 & ~~$-$23.62 & 3.10~~$\pm$~~0.20 &  14.07  \\
0347--121 & 0.188 & H & 18.28~~$\pm$~~0.15 &    17.72~~$\pm$~~0.01 &  $-$21.53 & ~~$-$22.30 & 1.25~~$\pm$~~0.05 &  ~3.93  \\
0350--371 & 0.165 & H & 18.03~~$\pm$~~0.15 &    17.08~~$\pm$~~0.01 &  $-$21.46 & ~~$-$22.60 & 1.70~~$\pm$~~0.07 &  ~4.81  \\
0414+009  & 0.287 & H & 16.08~~$\pm$~~0.05 &    17.49~~$\pm$~~0.02 &  $-$24.77 & ~~$-$23.71 & 4.70~~$\pm$~~0.50 &  20.31  \\
0419+194  & 0.512 & H & 19.53~~$\pm$~~0.17 &    21.05~~$\pm$~~0.15 &  $-$22.80 & ~~$-$22.22 & 0.40~~$\pm$~~0.07 &  ~2.47  \\
0502+675  & 0.314 & H & 17.33~~$\pm$~~0.10 &    18.86~~$\pm$~~0.09 &  $-$23.75 & ~~$-$22.62 & 0.60~~$\pm$~~0.07 &  ~2.76  \\
0506--039 & 0.304 & H & 18.73~~$\pm$~~0.15 &    18.35~~$\pm$~~0.01 &  $-$22.27 & ~~$-$23.03 & 0.60~~$\pm$~~0.05 &  ~7.19  \\
0521--365 & 0.055 & L & 15.28~~$\pm$~~0.10 &    14.60~~$\pm$~~0.01 &  $-$21.68 & ~~$-$22.41 & 2.80~~$\pm$~~0.07 &  ~3.00  \\
0548--322 & 0.069 & H & 16.93~~$\pm$~~0.10 &    14.62~~$\pm$~~0.01 &  $-$20.54 & ~~$-$22.92 & 7.05~~$\pm$~~0.15 &  ~9.31  \\
0607+710  & 0.267 & H & 18.23~~$\pm$~~0.10 &    17.83~~$\pm$~~0.02 &  $-$22.44 & ~~$-$23.17 & 2.40~~$\pm$~~0.12 &  ~9.85  \\
0706+591  & 0.125 & H & 17.53~~$\pm$~~0.07 &    15.94~~$\pm$~~0.01 &  $-$21.31 & ~~$-$23.04 & 3.05~~$\pm$~~0.07 &  ~6.84  \\
0735+178  & 0.424 & L & 16.58~~$\pm$~~0.07 & $>$20.44~~~~~~        &  $-$25.26 &  $>-$22.06 & ---               &  ~0.0~  \\
0737+744  & 0.315 & H & 17.88~~$\pm$~~0.15 &    18.01~~$\pm$~~0.08 &  $-$23.20 & ~~$-$23.48 & 2.10~~$\pm$~~0.45 &  ~9.68  \\
0806+524  & 0.138 & H & 15.98~~$\pm$~~0.02 &    16.62~~$\pm$~~0.01 &  $-$23.09 & ~~$-$22.61 & 1.45~~$\pm$~~0.03 &  ~3.53  \\
0823+033  & 0.506 & L & 17.78~~$\pm$~~0.11 & $>$20.18~~~~~~        &  $-$24.52 &  $>-$23.04 & ---               &  ~0.0~  \\
0828+493  & 0.548 & L & 18.93~~$\pm$~~0.12 &    20.26~~$\pm$~~0.10 &  $-$23.57 & ~~$-$23.31 & 0.65~~$\pm$~~0.10 &  ~4.16  \\
0829+046  & 0.180 & L & 15.88~~$\pm$~~0.07 &    16.94~~$\pm$~~0.04 &  $-$23.83 & ~~$-$22.98 & 4.30~~$\pm$~~0.75 &  13.06  \\
0851+202  & 0.306 & L & 14.99~~$\pm$~~0.06 & $>$18.53~~~~~~        &  $-$26.02 &  $>-$22.88 & ---               &  ~0.0~  \\
0927+500  & 0.188 & H & 17.48~~$\pm$~~0.30 &    17.62~~$\pm$~~0.05 &  $-$22.33 & ~~$-$22.40 & 2.00~~$\pm$~~0.45 &  ~6.29  \\
0954+658  & 0.367 & L & 16.08~~$\pm$~~0.06 & $>$19.60~~~~~~        &  $-$25.39 &  $>-$22.40 & ---               &  ~0.0~  \\
0958+210  & 0.344 & H & 21.48~~$\pm$~~0.40 &    18.93~~$\pm$~~0.01 &  $-$19.82 & ~~$-$22.84 & 0.82~~$\pm$~~0.04 &  ~4.01  \\
1011+496  & 0.2~~~& H & 15.88~~$\pm$~~0.05 &    17.30~~$\pm$~~0.02 &  $-$24.08 & ~~$-$22.89 & 1.80~~$\pm$~~0.12 &  ~5.94  \\
1028+511  & 0.361 & H & 16.48~~$\pm$~~0.10 &    18.55~~$\pm$~~0.08 &  $-$24.95 & ~~$-$23.40 & 1.80~~$\pm$~~0.35 &  ~9.08  \\
1104+384  & 0.031 & H & 13.78~~$\pm$~~0.08 &    13.29~~$\pm$~~0.02 &  $-$21.90 & ~~$-$22.41 & 3.95~~$\pm$~~0.05 &  ~2.45  \\
1133+161  & 0.460 & H & 20.28~~$\pm$~~0.18 &    19.83~~$\pm$~~0.04 &  $-$21.77 & ~~$-$22.99 & 1.55~~$\pm$~~0.23 &  ~9.05  \\
1136+704  & 0.045 & H & 16.15~~$\pm$~~0.04 &    14.45~~$\pm$~~0.02 &  $-$20.35 & ~~$-$22.10 & 3.10~~$\pm$~~0.02 &  ~2.75  \\
1212+078  & 0.136 & H & 16.38~~$\pm$~~0.10 &    16.02~~$\pm$~~0.01 &  $-$22.66 & ~~$-$23.17 & 3.40~~$\pm$~~0.10 &  ~8.19  \\
1215+303  & 0.130 & H & 14.55~~$\pm$~~0.01 &    15.99~~$\pm$~~0.01 &  $-$24.38 & ~~$-$23.08 & 8.35~~$\pm$~~0.20 &  19.36  \\
1218+304  & 0.182 & H & 15.68~~$\pm$~~0.10 &    17.12~~$\pm$~~0.03 &  $-$24.05 & ~~$-$22.82 & 2.60~~$\pm$~~0.30 &  ~7.97  \\
1221+245  & 0.218 & H & 16.89~~$\pm$~~0.05 &    18.63~~$\pm$~~0.06 &  $-$23.28 & ~~$-$21.79 & 1.25~~$\pm$~~0.25 &  ~4.42  \\
1229+643  & 0.164 & H & 18.03~~$\pm$~~0.30 &    16.38~~$\pm$~~0.01 &  $-$21.45 & ~~$-$23.29 & 2.00~~$\pm$~~0.07 &  ~5.63  \\
1248--296 & 0.370 & H & 18.83~~$\pm$~~0.08 &    18.87~~$\pm$~~0.02 &  $-$22.66 & ~~$-$23.15 & 1.10~~$\pm$~~0.05 &  ~5.63  \\
1255+244  & 0.141 & H & 17.08~~$\pm$~~0.05 &    16.72~~$\pm$~~0.01 &  $-$22.05 & ~~$-$22.57 & 2.50~~$\pm$~~0.05 &  ~6.21  \\
1402+041  & 0.340 & H & 16.38~~$\pm$~~0.01 & $>$19.38~~~~~~        &  $-$24.89 &  $>-$22.36 & ---               &  ~0.0~  \\
1407+595  & 0.495 & H & 18.84~~$\pm$~~0.05 &    19.04~~$\pm$~~0.05 &  $-$23.40 & ~~$-$24.08 & 1.75~~$\pm$~~0.38 &  10.63  \\
1418+546  & 0.152 & L & 15.68~~$\pm$~~0.06 &    16.10~~$\pm$~~0.02 &  $-$23.62 & ~~$-$23.37 & 3.65~~$\pm$~~0.11 &  ~9.65  \\
1426+428  & 0.129 & H & 17.38~~$\pm$~~0.20 &    16.14~~$\pm$~~0.01 &  $-$21.53 & ~~$-$22.92 & 2.25~~$\pm$~~0.08 &  ~5.18  \\
1440+122  & 0.162 & H & 16.93~~$\pm$~~0.12 &    16.71~~$\pm$~~0.02 &  $-$22.52 & ~~$-$22.93 & 3.90~~$\pm$~~0.25 &  10.87  \\
1458+224  & 0.235 & H & 15.78~~$\pm$~~0.08 &    17.80~~$\pm$~~0.05 &  $-$24.57 & ~~$-$22.83 & 3.20~~$\pm$~~0.80 &  11.96  \\
1514--241 & 0.049 & H & 14.48~~$\pm$~~0.12 &    14.45~~$\pm$~~0.01 &  $-$22.22 & ~~$-$22.30 & 3.70~~$\pm$~~0.10 &  ~3.55  \\
1534+014  & 0.312 & H & 19.08~~$\pm$~~0.15 &    18.16~~$\pm$~~0.02 &  $-$21.98 & ~~$-$23.31 & 2.00~~$\pm$~~0.10 &  ~9.16  \\
1704+604  & 0.280 & H & 21.08~~$\pm$~~0.35 &    18.69~~$\pm$~~0.01 &  $-$19.71 & ~~$-$22.45 & 0.85~~$\pm$~~0.03 &  ~3.61  \\
1728+502  & 0.055 & H & 16.43~~$\pm$~~0.10 &    15.49~~$\pm$~~0.02 &  $-$20.53 & ~~$-$21.52 & 3.15~~$\pm$~~0.05 &  ~3.37  \\
1749+096  & 0.320 & L & 16.88~~$\pm$~~0.05 &    18.82~~$\pm$~~0.10 &  $-$24.24 & ~~$-$22.73 & 3.00~~$\pm$~~0.80 &  13.97  \\
1757+703  & 0.407 & H & 18.43~~$\pm$~~0.14 &    19.58~~$\pm$~~0.25 &  $-$23.31 & ~~$-$22.77 & 0.85~~$\pm$~~0.50 &  ~4.62  \\
1807+698  & 0.051 & L & 14.95~~$\pm$~~0.25 &    13.87~~$\pm$~~0.02 &  $-$21.84 & ~~$-$22.97 & 2.10~~$\pm$~~0.10 &  ~2.09  \\
1853+671  & 0.212 & H & 19.48~~$\pm$~~0.10 &    18.19~~$\pm$~~0.01 &  $-$20.63 & ~~$-$22.17 & 1.50~~$\pm$~~0.08 &  ~5.18  \\
1959+650  & 0.048 & H & 15.38~~$\pm$~~0.10 &    14.92~~$\pm$~~0.02 &  $-$21.27 & ~~$-$21.78 & 5.10~~$\pm$~~0.10 &  ~4.80  \\
2005--489 & 0.071 & H & 12.73~~$\pm$~~0.01 &    14.52~~$\pm$~~0.01 &  $-$24.81 & ~~$-$23.09 & 5.65~~$\pm$~~0.08 &  ~7.65  \\
2007+777  & 0.342 & L & 18.03~~$\pm$~~0.10 &    19.03~~$\pm$~~0.10 &  $-$23.26 & ~~$-$22.73 & 3.30~~$\pm$~~0.90 &  16.07  \\
2143+070  & 0.237 & H & 18.21~~$\pm$~~0.11 &    17.89~~$\pm$~~0.02 &  $-$22.16 & ~~$-$22.76 & 2.10~~$\pm$~~0.15 &  ~7.90  \\
2200+420  & 0.069 & L & 13.58~~$\pm$~~0.05 &    15.37~~$\pm$~~0.02 &  $-$23.89 & ~~$-$22.17 & 4.80~~$\pm$~~0.40 &  ~6.33  \\
2201+044  & 0.027 & L & 17.18~~$\pm$~~0.05 &    13.74~~$\pm$~~0.01 &  $-$18.18 & ~~$-$21.64 & 6.78~~$\pm$~~0.08 &  ~3.68  \\
2254+074  & 0.190 & L & 16.94~~$\pm$~~0.12 &    16.61~~$\pm$~~0.02 &  $-$22.90 & ~~$-$23.44 & 4.90~~$\pm$~~0.35 &  15.54  \\
2326+174  & 0.213 & H & 17.63~~$\pm$~~0.11 &    17.56~~$\pm$~~0.03 &  $-$22.49 & ~~$-$22.81 & 1.80~~$\pm$~~0.15 &  ~6.24  \\
2344+514  & 0.044 & H & 16.83~~$\pm$~~0.05 &    14.01~~$\pm$~~0.01 &  $-$19.62 & ~~$-$22.49 & 5.93~~$\pm$~~0.02 &  ~5.14  \\
2356--309 & 0.165 & H & 17.28~~$\pm$~~0.13 &    17.21~~$\pm$~~0.02 &  $-$22.21 & ~~$-$22.47 & 1.85~~$\pm$~~0.10 &  ~5.23  \\
\enddata          
\tablenotetext{a}{SED type: H=HBL with $F_{1keV}/F_{5GHz} > 5.5$, and L=LBL, with $F_{1keV}/F_{5GHz} < 5.5$.}
\tablenotetext{b}{Apparent R-band magnitude of nucleus from best-fit PSF~+~de~Vaucouleurs model.}
\tablenotetext{c}{Apparent R-band magnitude of host galaxy from best-fit PSF~+~de~Vaucouleurs model.}
\tablenotetext{d}{Absolute R-band magnitude of nucleus, K corrected assuming $\alpha=1$.}
\tablenotetext{e}{Absolute R-band magnitude of host galaxy, K corrected assuming early-type spectrum.}
\tablenotetext{f}{De~Vaucouleurs effective radius in arcseconds.}
\tablenotetext{g}{De~Vaucouleurs effective radius in kiloparsecs, assuming $q_0 = 0$.}
\end{deluxetable}


\begin{table}
\begin{center}
\begin{tabular}{lcc}
Assumed    &Slope           &Intercept       \\
$z_{form}$ &(K-band mag/z)  &(K-band mag)    \\
\hline
2.8        &-1.37$\pm$0.30  &-24.97$\pm$0.11 \\
1.8        &-0.98$\pm$0.29  &-25.02$\pm$0.10 \\
1.3        &-0.65$\pm$0.28  &-25.04$\pm$0.10 \\
\hline
\end{tabular}
\end{center}
\caption{Regression analysis results for absolute luminosity evolution based 
on the range of evolution model fits to the Hubble diagram.
\label{regresstable}
}
\end{table}

\clearpage

\begin{figure}
\figurenum{1}
%
%
\caption{Central regions of STIS F28$\times$50LP images {\it right},
with contour plots {\it left}.\label{rawpics}}
\end{figure}

%
%
%
%
%
%
%


\begin{figure}
\figurenum{2}
\caption{
Two-dimensional and three-dimensional images of the composite stellar
PSF constructed from archival STIS F28$\times$50LP images of stars.
\label{stispsf}
}
\end{figure}


\begin{figure}
\figurenum{3}
\caption{Azimuthally averaged profiles of STIS images. 
  The {\it upper left} and {\it right} panels show the best-fit de~Vaucouleur
  + PSF models and exponential disk + PSF models respectively. 
{\it Middle} panels show zoomed-in profiles, inset with 1, 2 \&
3$\sigma$ $\chi^2$ contours projected on $m_{host}$--$r_e$ plane
(for cases where the host galaxy was confidently detected.) 
{\it Lower left} panels show the best-fit dual-component models and 
{\it lower right} panels show the best-fit PSF-only models.
For 1308+326, only the best PSF + exponential disk {\it upper} and
{\it lower left} and PSF-only {\it right} profiles are shown, while for
2240--260 the best PSF-only profile is shown.
\label{profiles}
}
\end{figure}

%
%
%
%
%
%
%
%
%
%
%
%
%
%
%
%

\begin{figure}
\figurenum{4}
\caption{BL Lac images with best-fit models subtracted ---
{\it top}: unsubtracted image; 
{\it middle}: PSF subracted, normalized according to best-fit de~Vaucouleurs model;
{\it bottom}: best PSF~+~de~Vaucouleurs model subtracted. For 1308+326
and 2240--260, the unsubtracted image ({\it top}) and best-fit PSF
subtraction assuming no host galaxy ({\it bottom}) are shown. The
arrow next to the object name indicates north and east.
In the case of 0235+164, we chose a gray scale to bring out the detail
in the surrounding structure, and so the marginal host galaxy is
difficult to make out.
\label{imagesub}}
\end{figure}

%
%
%
%
%
%
%
%

\begin{figure}
\figurenum{5}
\caption{
Hubble diagram for BL~Lac host galaxies, in F28$\times$50LP
magnitudes, corrected for galactic extinction. The {\it large points}
are the high-z sample: STIS targets from this study. The
{\it small points} are the low-z sample: WFPC2 snapshot targets with
$z<0.6$ from \citet{Urry}.
Also shown are the expected tracks of a non-evolving stellar population
with an early-type spectrum ($B-V=0.96$; {\it dashed line}), and a late-type
spectrum ($B-V=0.57$; {\it dotted line}).
\label{hubbleevol}
}
\end{figure}


\begin{figure}
\figurenum{6}
\caption{
 Hubble diagrams of high-z ({\it large points}) and low-z ({\it small
  points}) BL~Lac host galaxies. {\it Top} shows the best-fit
passively evolving model, with $z_{form}=1.8$ ({\it solid line}), and 
the $z_{form}=5$ passive model ({\it dashed line}).
{\it Middle} shows the best-fit non-evolving model which also fits 
the low-z colour constraint of $R-H>2$, with $age=5$~Gyrs 
({\it solid line}), as well as the $age=1$~Gyrs ({\it dashed line}) 
and $age=0.02$~Gyrs ({\it dotted line}) non-evolving models. This 
latter extremely young population is needed for a non-evolving 
population to fit the data. {\it Bottom} shows a dual population 
({\it solid line}), in which 98\% of the mass is evolving
passively from $z_{form}=5$, and 2\% is undergoing active star
formation with a constant age of 1~Gyr down to $z=0.6$, after which
it evolves passively. 
\label{hubblemodels}
}
\end{figure}


\begin{figure}
\figurenum{7}
\caption{
Hubble diagrams of high-z ({\it large points}) and low-z ({\it small
points}) BL~Lac host galaxies with evolutionary models
including dust and metallicity evolution. {\it Top left} shows 
a 0.8~Gyr non-evolving population with dust increasing linearly 
to an extreme optical depth of $\tau_V(0)=8$ between $z=0.8$ to 0.4 
({\it solid line}), giving a reasonable fit to the data. The 
{\it dashed line} shows dust increasing to $\tau_V(0)=2$, and the 
{\it dotted  line} shows the model with no dust, both yielding poor 
fits.
{\it Top right} shows a $z_{form}=5$ passive population with 
dust increasing from $\tau_V(0)=0$ to 10 by low redshift, 
producing a reasonable fit, while an increase to $\tau_V(0)=2$ 
({\it dashed line}) or a model with no dust ({\it dotted line}) produce
poor fits.
{\it Bottom left} shows the non-evolving population with 
metallicity increasing from 0.02~solar to 1~solar at $z=0.6$. 
The {\it solid line} produces a reasonable fit with dust also
increasing to $\tau_V(0)=5$ by low redshifts. An increase to 
$\tau_V(0)=2$ ({\it dashed line}) produces a marginal fit. The 
{\it dotted line} shows the model with no change in dust or 
metallicity.
{\it Bottom right} is the same as {\it bottom left}, but with a
$z_{form}=5$ passively evolving population, which requires the 
same dust and metallicity evolution to fit the data.
\label{dustmetalevol}
}
\end{figure}


\begin{figure}
\figurenum{8}
\caption{Absolute K-band magnitude versus redshift for high-z ({\it
  large dots}) and low-z ({\it small dots}) BL~Lac host galaxies. 
  These are K corrected assuming a
  redshift-dependent spectrum, determined using the evolution model
  which provided the best fit to the Hubble diagram (passive evolution with
  $z_{form}=1.8$; see \S\ref{lumevol}). Also
  plotted are the best-fit regression lines, derived using survival
  analysis to account for the upper limits. The {\it solid line} shows
  the best fit for $z_{form}=1.8$ K corrections, while the other lines
  represent the 3$\sigma$ upper and lower limits to the slope, determined assuming 
   $z_{form}=2.8$ K corrections ({\it dashed line}) and $z_{form}=1.3$ 
   K corrections ({\it dotted line}).
\label{mzregress}
}
\end{figure}

\end{document}